\documentclass[12pt]{article}
\usepackage{amsmath,cite,epsf,graphicx}

\def\beq{\begin{equation}}
\def\eeq{\end{equation}}
\def\bea{\begin{eqnarray}}
\def\eea{\end{eqnarray}}

\def \lsim
{\raisebox{-3pt}{$\>\stackrel{<}{\scriptstyle\sim}\>$}}

\begin{document}

\newenvironment{appendletterA}
{
  \typeout{ Starting Appendix \thesection }
  \setcounter{section}{0}
  \setcounter{equation}{0}
  \renewcommand{\theequation}{A\arabic{equation}}
 }{
  \typeout{Appendix done}
 }
\newenvironment{appendletterB}
 {
  \typeout{ Starting Appendix \thesection }
  \setcounter{equation}{0}
  \renewcommand{\theequation}{B\arabic{equation}}
 }{
  \typeout{Appendix done}
 }

%
%
%
%

\begin{titlepage}
\nopagebreak
{\flushright{
        \begin{minipage}{5cm}
          ROME1/1471-10 \\
        {\tt  [quant-ph] }
        \end{minipage}        }

}
\renewcommand{\thefootnote}{\fnsymbol{footnote}}
\vskip 2cm
\begin{center}
\boldmath
{\Large\bf 
Nonperturbative Analysis of a Quantum 
\\[7pt]
Mechanical Model for Unstable Particles}
\unboldmath
\vskip 1.cm
{\large  U.G.~Aglietti\footnote{Email: Ugo.Aglietti@roma1.infn.it},~~~
         P.M.~Santini\footnote{Email: paolo.santini@roma1.infn.it}, }
\vskip .2cm
{\it Dipartimento di Fisica, Universit\`a di Roma ``La Sapienza''  and
\vskip 0.1truecm
INFN, Sezione di Roma, I-00185 Rome, Italy} 
\end{center}
\vskip 0.7cm

\begin{abstract}

We present a detailed non-perturbative analysis of the 
time-evolution of a well-known quantum-mechanical system 
--- a particle between potential walls --- describing the 
decay of unstable states.
For sufficiently high barriers, corresponding to unstable 
particles with large lifetimes, we find an exponential decay 
for intermediate times, turning into an asymptotic power decay. 
We explicitly compute such power terms in time as a function 
of the coupling in the model. 
The same behavior is obtained with a repulsive as well as with 
an attractive potential, the latter case not being related to 
any tunnelling effect.

\vskip .4cm
{\it Key words}: quantum mechanics, unstable particle

\end{abstract}
\vfill
\end{titlepage}    

\setcounter{footnote}{0}

\section{Introduction and summary of results}

The decay of a particle originates from a coupling of its states, 
by virtue of some interaction, to a continuum of many-particle states, 
which constitute the decay products. 
It is necessary to have a coupling to an infinite number of 
states because, with a finite number of them, one only obtains an oscillatory 
behavior of particle amplitude with time, as it occurs for example with spin systems
\cite{feynman}.
Due to its intrinsic complexity, this process is usually treated 
in quantum field theory, as well as in quantum mechanics, in 
perturbation theory.
In lowest order, i.e. in the limit of vanishing coupling, 
one has a stable particle belonging to the spectrum of the free theory.
By switching on the interaction to the multi-particle continuum, the particle 
becomes unstable and disappears from the spectrum.
A common property of unstable particles is the exponential decay law,
which can be considered a law of nature:
\beq
\label{ed}
N(t) \, = \, N_0 \, e^{-t/\tau} \, ,
\eeq 
where $N(t)$ is the number of particles at time $t$, $N_0=N(0)$ is the initial number of particles,  
and $\tau$ is the mean lifetime.
Violation mechanisms of the exponential decay law have been proposed by various authors 
\cite{mt}, but have escaped experimental detection up to now.
Such effects manifest themselves in the form of power terms in the time, presumably 
with a small coefficient.
In this note we explicitly calculate such power effects in a simple quantum-mechanical
model, as a function of the interaction strength.
We consider a model similar to the one originally proposed to explain the $\alpha$-decay of nuclei: 
a particle initially confined between large and thick potential walls \cite{segre}.
Because of tunnelling effect, the wavefunction will filter
through the potential walls to an infinitely large region, so that the probability 
for the particle to remain in the original region will decrease with time down to zero. 
Even in our simple model, the time evolution cannot be computed in closed
analytic form and we derive rigorous asymptotic expansions for the wavefunction 
of the unstable state at large times; let us stress that the expansion parameter  
is $1/t$ and not the coupling of the interaction.

The main findings of our work are the followings.
For $|g| \ll 1$, where $g$ is the coupling, we find an exponential decay with a large
lifetime $\tau \approx 1/g^2$  in the time range
\beq
1 \, \ll \, t \, \lsim \,  \, \frac{\log(1/g)}{g^2} \, ,
\eeq
which turns into an asymptotic power decay $\approx g^4/t^3$ (for $|\psi|^2$).
By increasing $|g|$, the exponential time region reduces and eventually shrinks to zero.
That suggests that violations to the exponential decay law should be easier to
detect in strong coupling phenomena.
It is remarkable that a similar decay pattern is found for a repulsive as well as for
an attractive potential.
That was not expected on physical grounds, because there is no tunnelling effect  
for ``negative walls''. In other words, tunnelling is not the decay mechanism in
our model. 
In general, the decay properties seem to originate from the resonance characteristics 
of the system, which resembles a resonance cavity for small $|g|$ and occur both in 
the repulsive and in the attractive case. 

\section{The model}
\label{sec2}

The Hamiltonian operator of our model reads:
\beq
\hat{H} \, = \, - \, \frac{ \hbar^2 }{2 m} \frac{d^2}{d x^2} 
\, + \, \lambda \, \delta(x - L) \, ,
\eeq
where $m$ is the particle mass, $\lambda$ is a coupling constant and $x=L>0$ is the support
of the potential.
We assume that wavefunctions $\psi(x)$ are defined on the positive axis only,
$x \ge 0$, and that vanish at the origin: $\psi(0)=0$.
Formulae can be simplified by going to a proper dimensionless coordinate via
$x = L x' /\pi $ and rescaling the Hamiltonian as:
$\hat{H} = \hbar^2 \pi^2/(2 m L^2) \hat{H}'$.
The new Hamiltonian reads
\beq
\label{rescaled}
\hat{H'} \, = \, - \, \frac{d^2}{d x'^{\, 2} } \, + \, \frac{1}{\pi g}  \, \delta(x' - \pi),
\eeq
and contains the single real parameter $g \, = \, \hbar^2/(2 m L \lambda)$.
Let us omit primes from now on for simplicity's sake.

\section{The spectrum}

The eigenfunctions of the Hamiltonian operator $\hat{H}$ read:
\beq
\label{nondiviso}
\psi_k(x) \,\, \propto \,\, 
\Big( - \frac{i}{2} \, e^{i k x}   \, + \, \frac{i}{2} \, e^{- i k x} \Big) \, \theta(\pi - x)
\, + \,  
\Big(
a_k \, e^{i k x} \, + \, b_k \, e^{- i k x}
\Big)
\, \theta(x -  \pi)  \, ,
\eeq
and have eigenvalues $\epsilon_k = k^2$. $\theta(x)=1$ for $x>0$ and $0$ otherwise
is the step function and the coefficients $a_k$ and $b_k$ have the following
expressions:
\bea
a_k &=& - \, \frac{i}{2} \, - \, \frac{1}{4\pi g k} \left( 1 - e^{ - 2 i \pi k } \right) \, ;
\\
b_k &=& + \, \frac{i}{2} \, - \frac{1}{4 \pi g k} \left( 1 - e^{ + i 2 \pi k} \right) \, .
\eea
In general, $k$ is a complex quantum number.
For real $g$, $(a_k)^*=b_{k^*}$, implying that the zeros of the equation $a_k=0$
are complex conjugates of the ones of $b_k=0$. Since $b_{-k} = - a_k$, the zeroes
of $a_k=0$ are opposite to those of $b_k=0$. Furthermore, the eigenfunctions are 
odd functions of $k$, i.e. $\psi_{-k}(x) = - \psi_k(x)$, implying that it is necessary 
to consider only ``half'' of the complex $k$-plane.
However, in order to have real eigenvalues, as it should for the hermitian operator $\hat{H}$, 
$k$ must be either real or purely imaginary; the former case corresponds to non-normalizable
eigenstates with positive energies in the continuous spectrum, while the latter case corresponds 
to normalizable eigenfunctions with negative energies in the discrete spectrum, if any.
Since we have to deal both with normalizable and non-normalizable eigenfunctions, normalization
will be considered case by case in the next sections.

\subsection{Continuos spectrum}

The continuous spectrum is obtained for real $k$.
In this case $b_k = (a_k)^*$ and all the eigenfunctions are real.
Let us normalize them as:
\beq
\label{continuo}
\int_0^\infty \psi_{k'}^*(x) \, \psi_k(x) \, dx \, = \, \delta( k - k' ) \, ,
\eeq
where $\delta(q)$ is the Dirac $\delta$-function.
The normalization factor reads:
\beq
{\cal N}_k(g) \, = \, \frac{1}{ \sqrt{2\pi a_k b_k}  } 
\, = \,  
\sqrt{ \frac{2}{\pi} } \, \frac{1}{ \sqrt{ \, 1 \, + \, 1/(\pi g k) \sin 2 k \pi \, 
+ \, 1/(2 \pi^2 g^2 k^2) (1 - \cos 2 k \pi) \, } } \, .
\eeq
The final expression for the eigenfunctions therefore can be written as:
\beq
\label{diviso}
\psi_k(x) \, = \,
\frac{1}{\sqrt{2\pi}}
\left[
\left( 
- \frac{i}{2|a_k|} \, e^{i k x}   \, + \, \frac{i}{2|a_k|} \, e^{- i k x} 
\right) 
\, \theta(\pi - x)
\, + \,  
\left(
\frac{a_k}{|a_k|} \, e^{i k x} \, + \, \frac{{a_k}^*}{|a_k|} \, e^{- i k x}
\right)
\, \theta(x -  \pi)  
\right]
\, .
\eeq
Note that, because of continuum normalization, the amplitude of the eigenfunctions 
outside the wall is always $\mathcal{O}(1)$, no matter which values
are chosen for $k$ and $g$, while inside the cavity the amplitude has a 
non-trivial dependence on $k$ and $g$.
Note also that we can assume $k \, > \, 0$, implying that there is no energy degeneracy.

\subsection{Discrete spectrum}

Let us now consider the eigenfunctions with a purely imaginary quantum number $k$,
$k=ik_2$ with $k_2$ real, i.e. with the negative energy $\epsilon_{ik_2} = - {k_2}^2 < 0$ .
It holds:
\beq
\psi_{ik_2}(x) \,\, \propto \,\, 
\Big( - \frac{i}{2} \, e^{-k_2 x}   \, + \, \frac{i}{2} \, e^{+k_2 x} \Big) \, \theta(\pi - x)
\, + \,  
\Big(
a_{ik_2} \, e^{- k_2 x} \, + \, b_{ik_2} \, e^{+ k_2 x}
\Big)
\, \theta(x -  \pi)  \, .
\eeq
In order to obtain a normalizable state, the exponentially
growing term for $x \to + \infty$ must vanish, i.e. have zero coefficient 
--- quantization condition.
We can impose for example:
\beq
b_{ \, i k_2} \, = \, 0 ~~~ {\rm and} ~~~ k_2 > 0 .
\eeq
The equation is easy rewritten as:
\beq
\label{trascend}
e^{- 2 \pi k_2 } \, = \, 1 +  2 \pi k_2 g \, .
\eeq
The equation $e^{-s}=1+g s$, with $s\equiv 2\pi k_2$, has no solution for $g>0$, in agreement with 
physical intuition: there are no bound states with a repulsive
potential.
There is instead one non-trivial solution $k_2(g)$ for $-1<g<0$, again in
agreement with physical intuition:
\beq
\psi_{ik_2}(x) \, = \, 
C_{k_2} 
\left[
\theta(\pi - x) \left( e^{ \, k_2 \, x} -  e^{- k_2 \, x} \right)
\, + \, \theta(x -  \pi) \left( e^{ 2 \pi k_2 } - 1 \right) e^{- k_2 \, x}
\right]
\eeq
By normalizing to one, the overall coefficient reads:
\beq
C_{k_2} \, = \, \sqrt{ \, \frac{ k_2 }{ e^{2 \pi k_2  } - 1 - 2 \pi k_2 } ~ } \, .
\eeq
For $-1 \lsim g<0$, the trascendental equation for $k_2$ has the approximate solution 
\beq
k_2(g) \, \simeq \, \frac{g+1}{\pi} \, + \, \mathcal{O}\left[ (g+1)^2 \right] \, 
\eeq
while for a negative coupling of small size, $-1 \ll g < 0$,
\beq
k_2(g) \, = \, - \, \frac{1}{2\pi g} \Big[ 1 - e^{1/g} + \mathcal{O}\left( e^{2/g}  \right) \Big]  \, .
\eeq
Let us note that by imposing $a_{ik_2}=0$ and $k_2<0$, one obtains
the complex-conjugate zero $-k_2(g)$, in agreement with the symmetry property
of $a_k$ and $b_k$; the latter zero gives rise to the same eigenfunction.

\subsection{Normalizable eigenfunctions with complex energy}

In order to study the temporal evolution of general wavefunctions, 
it is convenient to consider normalizable eigenfunctions with a 
truly complex $k=k_1+ik_2$, i.e. with
the complex energy $\varepsilon = k_1^2-k_2^2+i 2 k_1 k_2$. 
The temporal evolution is controlled by the exponential factor
\beq
e^{-i\epsilon t} \, = \, e^{- i \left( k_1^2-k_2^2 \right) t + 2 k_1 k_2 t} \, ,
\eeq
implying decay with time for $k_1 k_2<0$. The eigenfunctions are of the form:
\bea
\psi_{k_1+ik_2}(x) &\propto& 
\Big[ - \frac{i}{2} \, e^{(ik_1-k_2) x}   \, + \, \frac{i}{2} \, e^{ (-ik_1+k_2) x} \Big] \, \theta(\pi - x)
\, +
\nonumber\\
&+&  
\Big[
a_{k_1+ik_2} \, e^{(i k_1- k_2) x} + b_{k_1+ik_2} \, e^{(-ik_1+ k_2) x}
\Big] \theta(x - \pi) \, .
\eea
As in the previous section, normalizable eigenfunctions are obtained by killing 
the exponentially growing terms for $x \to +\infty$.
By avoiding also an exponential growth with time, one obtains the relations:
\beq
b_{k_1 + i k_2} \, = \, 0 , ~~~~ k_1 < 0, ~~ k_2 > 0 .
\eeq
The above equation has a countable set of solutions $k^{(n)}(g)$
for any real $g$. For small $g$ ($n$ is a positive integer):
\beq
k^{(n)}(g) \, = \, - n + g n - g^2 n + i \pi g^2 n^2 + \mathcal{O}\left(g^3\right) \, .
\eeq
The distance of $k^{(n)}(g)$ from the real axis $\propto g^2$.
For small $g$, the eigenfunctions are confined inside the potential barrier, with 
an exponentially-decaying tail for $x>\pi$.

\section{Limiting cases}

The decay properties of our model become simpler in the limiting cases corresponding
to a large potential barrier, $|g|\ll 1$, and to a small potential barrier, $|g|\gg 1$.  

\subsection{Low Potential Barrier}

In the free limit, $g \to \pm \infty$, we have that $a_k \to -i/2$, $b_k \to i/2$ 
and the eigenfunctions reduce to sinusoidal waves in the whole positive axis: 
\beq
~~~~~~~~~~~~~~~~~
\psi_k(x) \, = \, \sqrt{ \frac{2}{\pi} } \, \sin ( k \, x ) ~~~~~~~~~~~~~ (g = \pm \infty) \, .
\eeq
There is clearly no discrete spectrum in this limit.

\subsection{High Potential Barrier}

Let us consider an eigenfunction in the continuous spectrum in a high potential barrier; 
formally we take
\beq
k \, \approx \, 1 ~~~ {\rm and} ~~~ |g| \, \ll \, 1 \, , 
\eeq
so that
\beq
|g| k \, \ll \, 1 \, .
\eeq
For most values of $k$,
\beq
\frac{1}{|a_k|} \, \approx \, |g| k \, \ll \, 1 \, ,
\eeq
i.e. the wavefunction amplitude is much smaller inside the potential wall than outside,
where it is always $\mathcal{O}(1)$.
Let us now determine the values of the quantum number $k$, if any,
for which the eigenfunctions have a large amplitude, i.e. an amplitude $\gg 1$, inside 
the potential barrier (i.e. for $0 < x < \pi$). 
By writing 
\beq
k \, = \, n \, + \, \xi^{(n)} \, ,
\eeq
with $n$ a non-zero integer, and imposing $1/|a_k|$ to be as large as possible 
\footnote{
There is no real $k$ which makes $b_k$ (or $a_k$) exactly vanishing for $g \ne 0$. 
As shown in the previous section, the expansion of the solution of the trascendental
equation $b_k = 0$ for small $g$ contains an imaginary part in second order in $g$.
},
we obtain:
\beq
k^{(n)}(g) \, = \, n \, - \, g n \, + \, \mathcal{O}\left( g^2 \right) \, .
\eeq
For these specific values, the amplitude inside 
the barrier ($0<x<\pi$) is rather large because
\beq
\frac{1}{\left|a_{k^{(n)}}\right|} \, = \, \mathcal{O}\left(\frac{1}{|g| n}\right) \, \gg \, 1 \, .
\eeq
In general, the amplitude is large inside the barrier
for values of $k$ inside intervals centered around $k^{(n)}(g)$ of size 
\footnote{
For $\xi_n = 0$ for example we obtain a pure sinusoidal wave in the whole 
positive axis, i.e. equal amplitudes inside and outside the barrier.
For these wavelenghts, there is actually no effect of the potential (transparency).
}
\beq
\delta k^{(n)}(g) \, \approx \, |g| n \, .
\eeq
For these ranges of $k$, the region between the potential walls resembles
a resonant cavity. 
On the contrary, for values of $k$ outside the intervals $(n-|g|n n,n+|g|n)$, 
the wave amplitude is small inside the barrier, i.e. it is $\ll 1$.
Finally, there is a large phase shift  when the inside amplitude becomes 
$\ll 1$ from $\gg 1$ and vice-versa \cite{flugge}.

Let us stress that a similar resonant behavior is found for $g>0$ as well as for $g<0$,
as it stems from the above formulae.
Since
\beq
\lambda^{(n)}(g) 
\, = \, \frac{2\pi}{k^{(n)}(g)} 
\, = \, \frac{2\pi}{n} \Big[ 1 + g + \mathcal{O}\left(g^2\right)  \Big] \, ,
\eeq
the only difference is that for $g>0$ the ``resonant'' wavelengths are slightly above
the natural frequencies $2\pi/n$ of the cavity while in the attractive case they are
slightly below.

\subsubsection{Infinite Barrier}

Let us now consider the limit $g \to 0$ at fixed $n$,
in which the potential wall becomes infinitely high. We find that
\beq
k^{(n)} \, \to \, n \, , ~~~~~ \delta k^{(n)} \, \to \, 0 \, , ~~~~~ a_{k^{(n)}} \, \to \, 0 \, .
\eeq
Then, for the discrete momenta $k^{(n)}=n$, the wavefunction is completely 
inside the ``cavity'', while for all the remaning values, the wavefunction is 
completely outside it.
The values $k^{(n)} = n$ above correspond to the quantised (allowed)
momenta of a particle in a one-dimensional box of length $l = \pi$, implying that the potential
wall becomes impenetrable in the limit $g \to 0$, in agreement with physical intuition:
there is no coupling of the cavity with the outside \cite{flugge}.
Outside the cavity, we have instead a continuous spectrum 
of eigenfunctions labelled with $k$, defined for $x \ge \pi$ and vanishing in $x=\pi$.
\footnote{
Since $k$ is a continuous variable, removing integer values (a zero measure set) has no 
influence on the spectrum.} 
We may say that, in the limit $g \to 0$, the system decomposes into two 
non-interacting sub-systems, representing stable particles and a continuum of
multi-particle states.
It is a remarkable formal fact that in the limit $g \to 0$ the eigenfunctions
corresponding to the eigenvalues $k^{(n)}$ become normalizable:
since $a_{k^{(n)}} = 0$, eq.~(\ref{diviso}) becomes meaningless and one has to go back to 
eq.~(\ref{nondiviso}) and to impose discrete-state normalization --- typically normalization 
to one.

Let us now consider the properties of the discrete spectrum of our model for $g\to 0$.
For $g \to 0^+$ there are no bound states, while for $g \to 0^-$
there is always one bound state which gets progressively more localized around
the potential support $x=\pi$. Heuristically, this discontinuous behaviour of 
the discrete spectrum around $g=0$ suggests that this point a singular one.
\begin{figure}[ht]
\begin{center}
\includegraphics[width=0.5\textwidth]{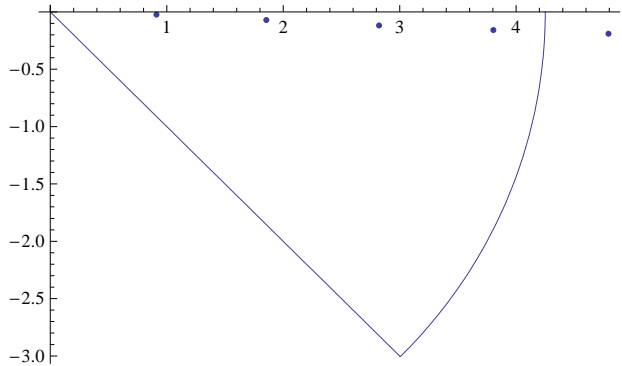}
\footnotesize\caption{
\label{figura1}
\it Integration contour in $k$-plane for $g=0.1$.}
\end{center}
\end{figure}

\section{Time evolution for the unstable state}
\label{sec3}

This is the central section of the paper, in which we compute the time evolution
of unstable states by means of non-perturbative analytic techniques.
The first problem is therefore that of defining an unstable particle
and the second one to find the analog of such a state in our model.
In physical terms, an unstable particle (or resonance) is related to
the enhancement of any cross section producing that particle in the $s$-channel
in a narrow band of $P^2 \approx M^2$, where $P_{\mu}$ is the total 4-momentum
and $M$ the particle mass.
An unstable particle is therefore a quantum (i.e. coherent) state prepared at 
$t = 0$, which is normalizable and has a non-trivial time evolution,
not being an eigenstate of the system
\footnote{
A general definition of an unstable particle via S-matrix elements
constructed by means of wave-packets has been provided in \cite{mt}
(for general background see for example \cite{newton}).
}.
In general, there is some freedom in the definition of such states.
We try to keep as close a connection as possible with the procedure
followed in perturbation theory: we take as unstable state an exact eigenstate 
of the non-interacting system ($g=0$), which is no more an exact eigenstate of 
the interacting system ($g \ne 0$).
Let us then consider the evolution of a wave-function given at the initial time 
$t \, = \, 0$ by
\beq
\label{prepare}
\psi^{(l)}(x, \, t = 0) \, = \, \sqrt{\frac{2}{\pi}} \, \sin \left( l \, x \right) \, \theta(\pi - x) \, , 
\eeq
where $l$ is a positive integer.
Note that, unlike the eigenfunctions $\psi_k(x)$ considered above, 
$\psi^{(l)}(x,0)$ is a normalized state:
\beq
\int_0^{\infty} \left| \psi^{(l)}(x, 0) \right|^2 \, dx \, = \, 1 \, . 
\eeq
Let us remark that any continuous wavefunction with support in the interval $(0,\pi)$
--- representing the more general state for an unstable particle ---
can be expressed as superposition of the above wavefunctions.
As discussed in the previous section, for $g=0$ the wavefunction above 
cannot decay being an eigenfunction and therefore we expect it to represent
a slowly decaying state for $0 < |g| \ll 1$.
The decay products corresponds to non-normalizable wavefunctions in our model
of the form
\beq
\label{products}
\phi_k(x,t) \, \propto \, \theta(x - \pi) \, \sin \big[ k (x-\pi) \big] \, ,
\eeq
vanishing for $x \le \pi$.
For $g \ne 0$, also the wavefunctions above are not eigenfunctions;
in physical terms, they can excite modes inside the cavity.
The wavefunctions in eq.~(\ref{prepare}) exactly vanish outside the barrier,
implying there are no ``decay products'' at $t = 0$.

\noindent
To compute the time evolution, namely
\beq
\psi^{(l)}(x, \, t) \, = \, e^{-i \hat{H} t } \, \psi^{(l)}(x, \, 0) \, ,
\eeq
the most convenient technique is to expand the state above 
into eigenfunctions of the Hamiltonian which, as well known, have trivial time evolution:
\beq
\psi^{(l)}(x, \, 0) \, = \, \int_0^\infty \varphi_k^{(l)} \, \psi_k(x) \, dk \, ,
\eeq
where:
\beq
\varphi^{(l)}_k \, =  \, (-1)^l \, l \, \frac{2}{\pi} \, N_k(g)  \, \frac{\sin k \pi}{ k^2 \, - \, l^2 } \, ,
\eeq
satisfying $\varphi^{(l)}_{-k} = - \varphi_k^{(l)}$, with
\beq
N_k^2(g) \, = \, 
\frac{1}{4 a_k b_k}
\, = \, 
\frac{1}{ 1 \, + \, 1/(\pi g k) \sin 2 k \pi \, 
+ \, 1/(2 \pi^2 g^2 k^2) (1 - \cos 2 k \pi) } \, .
\eeq
We have considered for simplicity's sake only the repulsive case $g>0$.
Our wavefunction at time $t$ is therefore given by:
\footnote{
Since the integrand is an even function of $k$ because both $\varphi_k^{(l)}$ and $\psi_k(x)$ are odd,
one can extend the integral over all $k$'s as
\beq
\int_0^\infty dk \, \to \,  \frac{1}{2} \int_{-\infty}^\infty dk \, .
\eeq
}
\beq
\label{wf}
\psi^{(l)}(x,\, t) \, = \, \int_0^\infty \varphi^{(l)}_k \, \psi_k(x) \, e^{ - i \, k^2 \, t } \, dk \, .
\eeq

\subsection{Asymptotic Expansion of Wave Function}

The spectral representation in eigenfunctions of the unstable state at time $t$ has
the explicit expression:
\beq
\label{psi}
~~~~~~~
\psi^{(l)}(x,t) \, = \, 
\left(\frac{2}{\pi}\right)^{3/2} \int\limits_{0}^{\infty} p^{(l)}(k; x, g) \, e^{-ik^2t} \, dk , 
~~~~~~~ 0 \, \le \, x \, \le \, \pi \, , ~~~ g \, > \, 0 \, ,
\eeq 
where
\beq
p^{(l)}(k;x,g) \, = \, (-1)^l l \frac{\sin k \pi}{ k^2 - l^2 } 
\, \frac{1}{ 1 + 1/(\pi g k) \sin 2 k \pi + 1/(2 \pi^2 g^2 k^2) ( 1 - \cos 2 k \pi ) }
\, \sin k x \, .
\eeq
To obtain rigorous analytic formulae, we expand the integral for large $t$. 
The steepest descent method suggests to replace the integral on the r.h.s. of eq.~(\ref{psi}) 
by the integral over the steepest descent ray ($0,\infty e^{-i\pi/4}$), on which 
the fast oscillation of the integrand is absent. This is achieved by
computing the above integral over the sequence of closed contours $\gamma_n$ 
containing the segment on the real axis $(0,n+\alpha_n)$, 
the ray $((n+\alpha_n)e^{-i\pi/4},0)$ and a circular arc $c_n$ connecting the 
endpoints of the two segments (see fig.~1).
The parameter $\alpha_n$ ($0\le \alpha_n \le 1/2)$ is chosen is such a way that $c_n$ passes at the
maximal possible distance between the poles in the fourth quadrant of the $k$-plane.
Therefore the state $\psi^{(l)}(x,t)$ is decomposed in a natural way into the sum of two quite
different contributions:
\beq
\label{residue}
\psi^{(l)}(x,t) \, = \, \psi^{(l)}_{pow}(x,t) \, + \, \psi^{(l)}_{exp}(x,t) \, ,
\eeq
where
\bea
\psi^{(l)}_{pow}(x,t) &\equiv& 
e^{-i\pi/4}\left(\frac{2}{\pi}\right)^{3/2} \, \int\limits_{0}^{\infty} p^{(l)}\left( k \, e^{-i\pi/4}; x,g \right) 
\, e^{-k^2 t} \, dk \, ; 
\\
\label{residues}
\psi^{(l)}_{exp}(x,t) &\equiv& - 2 \pi i\left(\frac{2}{\pi}\right)^{3/2}\sum_{n=1}^{\infty} {\rm Res} 
\left[ p^{(l)}(k; x,g) \, e^{- i k^2 t} , \, k^{(n)} \right] \, .
\eea    
In general, the contribution $\psi^{(l)}_{pow}(x,t)$ exhibits a power decay as $t \gg 1$, 
while the contribution $\psi^{(l)}_{exp}(x,t)$, coming from the residues in 
eq.~(\ref{residues}), exhibits an exponential decay.
Let us consider the above contributions in turn:
\begin{enumerate}
\item
the integral $\psi^{(l)}_{pow}(x,t)$ is over the ray ($0,\, \infty \, e^{-i\pi/4}$)
and for large $t \gg1$ takes the dominant contribution from a neighborhood of $k = 0$, 
where the integrand is analytic and can therefore be expanded in powers of $k$:
\beq
\label{pj}
p^{(l)}(k;x,g) \, = \, \frac{g^2}{(1+g)^2}\sum\limits_{j=1}^{\infty}p^{(l)}_j(x,g) \, k^{2j} \, .
\eeq
The first few coefficients explicitly read:
\bea
p_1^{(l)}(x,g) &=& \frac{ (-1)^{l+1} }{l} \, \pi x \, ;
\\
p_2^{(l)}(x,g) &=& \frac{ (-1)^{l+1} }{l} \pi x
\left[
\frac{1}{l^2} + \frac{\pi^2}{6} + \frac{2}{3} \frac{\pi^2 g}{1+g} 
- \frac{\pi^2 g^2}{(1+g)^2} - \frac{x^2}{6}
\right] \, .
\eea
Replacing this series into the integral and performing the change of variable $\nu = k^2 \, t$, 
one obtains the following asymptotic expansion:
\bea
\psi^{(l)}_{pow}(x,t)
& \approx & \frac{\sqrt{2}}{\pi^{3/2}}\frac{e^{-i\pi/4}g^2}{(1+g)^2}
\sum\limits_{j=1}^{\infty}\frac{(-i)^j \, p_j^{(l)}(x,v)}{t^{j+\frac{1}{2}}}
\int\limits_{0}^{\infty} d\nu \, \nu^{j-\frac{1}{2}} \, e^{-\nu}
\\
&=&
\frac{\sqrt{2}}{\pi}\frac{e^{-i\pi/4}g^2}{(1+g)^2}\sum\limits_{j=1}^{\infty}\frac{(-i)^j(2j-1)!!}{2^j}
\frac{p_j^{(l)}(x,g)}{t^{j+\frac{1}{2}}} , ~~ 0 \le x \le \pi , ~ t \gg 1 ,
\eea
whose first few terms read:
\bea
\psi^{(l)}_{pow}(x,t) &\approx&
\frac{ e^{i\pi/4} }{\sqrt{2}} \frac{ (-1)^l }{l} \frac{g^2}{(1+g)^2} \frac{x}{ t^{3/2} }
\Bigg\{ 
1 \, - \,  
\frac{3 i}{2t} 
\Bigg[
\frac{1}{l^2} + \frac{\pi^2}{6} + \frac{2}{3} \frac{\pi^2 g}{1+g} \, +
\nonumber\\
&&~~~~~~~~~~~~~~~~~~~~~~~~~~~~~~~~~~~
- \frac{\pi^2 g^2}{(1+g)^2} - \, \frac{x^2}{6}
\Bigg]
+ \mathcal{O} \left( \frac{1}{t^2} \right)
\Bigg\} \, . 
\eea
Let us make a few remarks.
The above asymptotic expansion is uniformely valid for all $g \ge 0$, since 
the coefficients $p_j^{(l)}(x,g)$ are uniformely bounded in that region 
(see eq.~(\ref{pj})).
The exponent $3/2$ controlling the power decay, $\psi \approx 1/t^{3/2}$,
does not depend on $l$ and $g$; 
\item
the simple poles $k^{(n)}(g)$ of the integrand in $\psi_{exp}^{(l)}(x,t)$
are the simple zeroes of the trascendental equation
\beq
\label{poles1}
4a_k b_k \, = \, 1 + \frac{ \sin 2 k  \pi  }{ \pi g k } 
+ \frac{ 1 - \cos 2 k \pi }{ 2 \pi^2 g^2 k^2  } \, = \, 0 \, ,
\eeq 
constrained by the conditions:
\beq\label{poles2}
\mathrm{Re} \, k^{(n)} \, > \, |\mathrm{Im} \, k^{(n)}| \, , ~~~ \mathrm{Im} \, k^{(n)} \, < \, 0 \, .
\eeq
It is easy to prove that all the poles
$k^{(n)}(g)$ satisfy all of the above conditions for $g \ll 1$.
The second condition in (\ref{poles2}) is always satisfied.
In general, the poles leave the fourth quadrant for very large values of $|g|$, where the
unstable-particle description is irrelevant. 
\begin{figure}[ht]
\begin{center}
\includegraphics[width=0.5\textwidth]{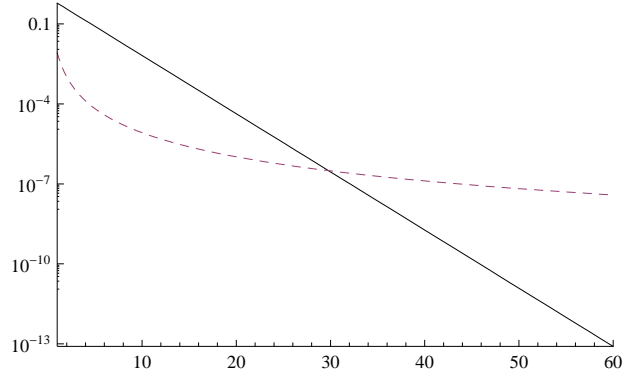}
\footnotesize\caption{
\label{figura2}
\it Time evolution of the modulus square of the exponential contribution 
(continuous line) and power contribution (dashed line) to the wavefunction 
of the fundamental state $l=1$ (see eq.(\ref{prepare})) for $g=0.2$, 
integrated in the interval $(0,\pi)$. 
The scale on the vertical axis is logarithmic.}
\end{center}
\end{figure}
The unstable states (or resonances) have pole contributions for $g \ll 1$ of the form:
\beq
\label{nuova}
\psi_{exp}^{(l)}(x,t) \, = \, \psi_{pole}^{(l)}(x,t)
\, + \, g \sum_{n \, \ne \, l}^{1,\, \infty} \, c_{ \, l,\,n} \, \psi_{pole}^{(n)}(x,t) \, ,
\eeq
where
\bea
c_{\, l,\, n} &\equiv& (-1)^{l+n} \frac{  2 \, l \, n }{ l^2 - n^2 } \, ;
\\
\psi_{pole}^{(n)}(x,t) &\equiv& \sqrt{ \frac{2}{\pi} } \, Z^{(n)}(g) \,
\sin \left[ k^{(n)}(g) x \right] \, e^{ - i \, \omega^{(n)}(g) \, t - 1/2 \, \Gamma^{(n)}(g) \, t } \, ,
\eea
with
\bea
Z^{(n)}(g) &=& 1 - \frac{g}{2} \, + \mathrm{O}\left( g^2 \right) \, ;
\\
\omega^{(n)}(g) &\equiv& + \, \left( \mathrm{Re} \, k^{(n)} \right)^2 \, - \, \left( \mathrm{Im} \, k^{(n)} \right)^2 
\nonumber\\
&=& n^2 \left[ 1 - 2g \, + \mathrm{O}\left( g^2 \right) \right] \, ;
\\
\Gamma^{(n)}(g) &\equiv& - \, 4 \, { \mathrm{Re} \, k^{(n)} } \, { \mathrm{Im} \, k^{(n)} }
\nonumber\\
\label{width2e3}
&=&  4 \pi n^3 g^2 
\left[
1 \, - \, 4 g \, + \, {\mathcal O}\left( g^2 \right)
\right] \, .
\eea
\end{enumerate}

\section{Discussion}
\label{sec4}

Let us consider the asymptotic expansion for $t \gg 1$ of the first 
resonance ($l=1$) by neglecting the contributions of the higher-order poles 
($n>1$):
\bea
\label{mainres}
\psi^{(1)}(x,t) &\approx&
- \frac{ e^{i\pi/4} }{\sqrt{2}} \frac{g^2}{ (1+g)^2 } \frac{ x }{ t^{3/2} }
\left\{ 
1 \, - \, 
\frac{3 i}{2t} \, 
\Bigg[
1 + \frac{\pi^2}{6} + \frac{2}{3} \frac{\pi^2 g}{1+g} \,
- \frac{\pi^2 g^2}{(1+g)^2} - \, \frac{x^2}{6}
\Bigg]
+ \mathcal{O}\left( \frac{1}{t^2} \right)
\right\} +
\nonumber\\
&+& \sqrt{ \frac{2}{\pi} } 
\left( 1 - \frac{g}{2} \right)
\, \sin 
\left[ 
\left( 1 - g \right) x 
\right] 
\, \exp
\left[
 - i  \left( 1 - 2 g \right)  t 
\, - \, 2 \pi g^2 t
\right] \, .
\eea
A few comments are in order:
\begin{enumerate}
\item
in the limit $g \to 0$, the power term disappears from the r.h.s. of eq.~(\ref{mainres})
and the exponential term approaches the fundamental eigenfunction of a particle with mass $m=1/2$ 
in a box of length $l=\pi$:
\beq
\label{stable}
\Psi^{(1)}(x,t) \, \to \, \sqrt{ \frac{2}{\pi} } \, \sin x \, e^{ - i  \, t } \, , ~~~~ g \to 0 \, .
\eeq
That is in complete agreement with physical intuition, as already discussed;
\item
it is clear that, for sufficiently long times, the decay law will be dominated by the power term. 
However, since the power term has a small coefficient, suppressed as $g^2 \ll 1$
for small  $g$, while the exponential term has a coefficient
of order one and it remains $\mathcal{O}(1)$ as long as $t \lsim 1/g^2$,
the exponential term dominates over the power term for a large temporal region for $g \ll 1$.
In other words, if $g$ is small and $t$ is large, but such that 
\beq
1 \, \ll \, t \, \lsim \,  \, \frac{\log(1/g)}{g^2} \, ,
\eeq
there is a long transient in which the exponential term 
prevails on the power term (see fig.~2).
Since the signal rapidly decays with time, the transient region may actually be
the one measurable one.
\end{enumerate}

\vskip 0.5truecm

\centerline{\bf Acknowledgements}

\vskip 0.5truecm
\noindent
One of us (U.G.A.) would like to thank D.~Anselmi and M.~Testa for discussions.

\vskip 0.6truecm

\centerline{\bf NOTE ADDED}

\vskip 0.4truecm

\begin{figure}[ht]
\begin{center}
\includegraphics[width=0.5\textwidth]{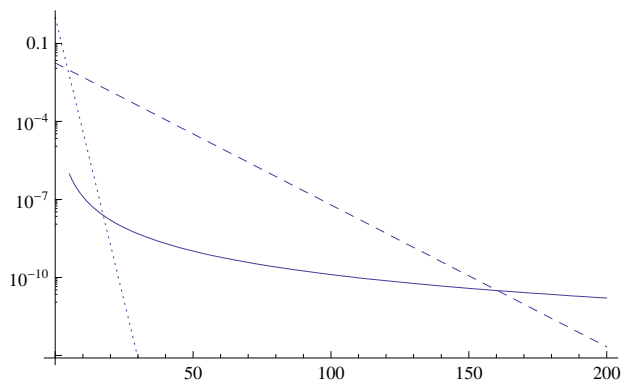}
\footnotesize
\caption{
\label{figura3}
\it Time evolution of the contributions to the $l=2$, i.e. first excited, state
for $g=0.1$.
Dotted line: second pole contribution;
Dashed line: first pole contribution;
Continuous line: power contribution.}
\end{center}
\end{figure}

\noindent
After the first version of this note was put on the archive, 
references \cite{general} and \cite{Winter:1961zz} were brought 
to our attention\footnote{We wish to thank Dr. R.~Rosenfelder
for pointing out \cite{Winter:1961zz} to us.}.
While the first paper deals with general properties of the exponential 
region, the second one treats the time evolution
with the saddle point method of the same model as we do.
We are in complete agreement with \cite{Winter:1961zz} as far
as the asymptotic power behavior in time is concerned, while we are in
disagreement with the exponential behavior.
More specifically, the first term on the r.h.s. of our eq.(\ref{nuova}),
i.e. the diagonal one $n=l$, is in
agreement with the r.h.s. of eq.(2a) in \cite{Winter:1961zz} 
($l$ labels the initial state and $n$ the pole).
In \cite{Winter:1961zz} however, the non-diagonal pole contributions
$n \ne l$, present in eq.(\ref{nuova}), are not included. These terms have a 
coefficient suppressed by a power of $g\ll 1$ compared to the
diagonal one, but have a slower exponential decay for
$n<l$ ($\Gamma^{(n)}\propto n^3$, see eq.(\ref{width2e3})), 
and therefore dominate at intermediate times (i.e. before
power-effects take over). As shown in fig.\ref{figura3}, there is indeed a large
temporal region where the non-diagonal contribution from the first pole,
\beq
\frac{16}{9} g^2 \int_0^{\pi}|\psi_{pole}^{(1)}(x,t)|^2 dx \, ,
\eeq
dominates over that of the second pole, 
\beq
\int_0^{\pi}|\psi_{pole}^{(2)}(x,t)|^2dx \, ,
\eeq
in the temporal evolution of the first excited state, $l=2$.
Neglecting the non-diagonal contributions
is therefore a reasonable approximation only for the time-evolution of 
the lowest-lying state $l=1$. 
More details will be given in a forthcoming publication \cite{seconda}.


\begin{thebibliography}{99}

\bibitem{feynman}
R.~P.~Feynman, {\it La Fisica di Feynman (The Feynman Lectures in Physics)},
Masson Italia Editori, Milano (1985), vol.~3.

\bibitem{degasperis}
A.~Degasperis and L.~Fonda, {\it Does the life-time of an Unstable Particles Depend
on the Measuring Apparatus?}, Il Nuovo Cimento vol.~21, n.~3 (1974).

\bibitem{mt}
L.~Maiani and M.~Testa, {\it Unstables Systems in Relativistic Quantum Field Theory},
Ann.~of~Phys. vol.~263, n.~2, pag.~353 (1998). 

\bibitem{segre}
See for example: E.~Segre, {\it Nuclei e Particelle}, Zanichelli Ed.~(1982), chap.~7.

\bibitem{flugge}  
S.~Flugge, {\it Practical Quantum Mechanics}, Springer-Verlag, Berlin (1994), 
problem n.~27.

\bibitem{newton}
R.~Newton, {\it Scattering of Waves and Particles}, Dover Publications, Inc.
Mineola, New-York (2004);
M.~Goldberger and K.~Watson, {\it Collision Theory}, Dover Publications, Inc.
Mineola, New-York (2002).

\bibitem{general}
N.~Hatano et al, Prog.~Theor.~Phys. Vol. 119 n.2 pag. 187 (2008) and
references therein. 

\bibitem{Winter:1961zz}
R.~G.~Winter,
Phys.\ Rev.\  {\bf 123}, 1503 (1961).

\bibitem{seconda}
U.G.~Aglietti and P.M.~Santini, in preparation.

\end{thebibliography}
\end{document}